\documentclass[twocolumn]{IEEEtran}  \newdimen\figwidth  \figwidth=8.3cm

\usepackage{cite}
\usepackage[english]{babel}
\usepackage{amsmath}
\usepackage{color}
\usepackage{graphicx}
\usepackage{pbox}
\usepackage{booktabs}
\usepackage[center]{caption}
\usepackage{subcaption}
\usepackage{mdframed}
\usepackage{balance} 
\usepackage{epsfig}   
\usepackage{amssymb} 

\begin{document}
\title{A New Relation Between\\ Energy Efficiency and Spectral
Efficiency\\ in Wireless Communications Systems}
\author{
\IEEEauthorblockN{\large Lokman Sboui$^{\text{1}}$, Zouheir Rezki$^{\text{2}}$, Ahmed Sultan$^{\text{3}}$, and Mohamed-Slim Alouini$^{\text{3}}$}\\[.2cm]
\IEEEauthorblockA{\small
$^{\text{1}}$\'Ecole de Technologie Sup\'erieure (\'ETS), Montreal, Canada. \\
$^{\text{2}}$University of Idaho, Moscow, ID, USA, \\
$^{\text{3}}$King Abdullah University of Science and Technology (KAUST), Thuwal, Makkah Province, Saudi Arabia.\\\vspace{2mm}
\small lokman.sboui@lacime.etsmtl.ca, zrezki@uidaho.edu, ahmed.salem@kaust.edu.sa, slim.alouini@kaust.edu.sa\\}
}
\maketitle
\begin{abstract}
When designing wireless communication systems (WCS), spectral efficiency (SE) has been the main design performance metric. Recently, energy efficiency (EE) is attracting a huge interest due to the massive deployment of power limited WCS such as IoT devices, and stringent environmental concerns. For this reason, many works in the literature focused on optimizing the EE and highlighted the EE-SE relationship as a trade-off (meaning that increasing one decreases the other). In this article, after introducing the EE metric, we highlight a new perspective of the EE-SE relationship based on energy-efficient power control. In particular, we give insights about the EE-based performance of various transmission technologies and its impact on 5G future design. Via numerical results, we show that the corresponding power scheme allows an increase of both the SE as the EE with no trade-off. Finally, we present relevant open research problems.
\end{abstract}
%
%
\section{Introduction}

During the last decade, the main objective of wireless communications systems (WCS) is to ensure reliable content delivery and to enhance the quality of service. Hence, efforts in industry and academia have been made to improve the spectral efficiency (SE) and obtain higher throughput based on the Shannon capacity that describes the maximum amount of data which can be transmitted over a wireless channel.
The focus on maximizing the SE resulted in developing many technologies such as orthogonal frequency-division multiplexing (OFDM), multiple-input and multiple-output (MIMO) and relaying techniques, to cite only a few.

Currently, the applications of the fifth generation (5G) WCS are expected to be very diverse; ranging from environmental monitoring to public safety. Consequently, the energy efficiency (EE)  is becoming an important performance metric when designing such systems. One of the 5G requirements is to increase the energy efficiency by 100 times, as indicated by the International Telecommunications Union (ITU) in its report~\cite{ITUreport2015}.
Such a metric is extremely important and crucial in multiple scenarios.
Firstly, for many battery-powered devices like the WSN and mobile devices, operation lifetime is a key parameter and is required to be as long as possible to ensure continuous and long-lasting services.
Secondly, for the cellular operators, the energy consumption constitutes an important part of the operating expenditure (OPEX). Consequently, reducing the energy consumption while conserving the same QoS is important to reduce the OPEX and enhance the profit given the increasing energy prices. The work in~\cite{Xu2015} describes how smart grids were used in cellular networks in order to improve the EE and reduce the related cost.
Thirdly, since most of the current electrical energy is generated via fossil processes, the CO$_2$ emission is another concern and a stringent constraint that must be respected by the service providers.
Finally, it is also important to evaluate the energy consumption of critical and low-latency applications by analyzing and maximizing the corresponding EE.

Previous works indicated that maximizing the EE and SE are conflicting objectives, that is, in order to increase the EE, we need to reduce the SE~\cite{Li2011}. In fact, since the focus is on maximizing the SE, many power allocations schemes, such as water-filling, were developed to improve the SE~\cite{Cover2006}. Hence, the corresponding  EE was compromised when adopting such power control schemes. Therefore, several green projects have been introduced to cope with such a compromise, e.g.,~\cite{Han2011}. In some other works, e.g.,~\cite{Guo2014}, a joint energy and spectrum cooperation was presented to avoid this trade-off. Consequently, the resource allocation schemes using the EE as an objective function are becoming popular.
In fact, in the ITU report~\cite{ITUreport2015}, it was mentioned that in order to improve the EE, the transmit power and the circuit power should be reduced.
However, this approach did not seem to be promising since it undermines the SE. Indeed, it has been shown in~\cite{Li2011} that there is an EE-SE trade-off, in the sense that one cannot maximize both figures of merits at the same time, and in order to maximize one, the other needs to be reduced.

In this paper, and for the first time, we show that the trade-off highlighted by previous works is actually not fundamental. In particular, we present a new EE-SE relationship where both metrics concurrently increase.
In addition, we show that there exists a power control scheme that maximizes EE while having an increasing SE. In particular, in~\cite{Sboui2015}, it has been shown that a power control scheme that maximizes the EE also exhibits an increasing EE as a function of the SE.
In the rest of this article, we study different cases and scenarios;  single-input-single-output (SISO), OFDM, and MIMO that cover the majority of current WCS. We also present some of the open research problems related to the EE of WCS.

\section{Background on the EE Metric}
The concept of EE for WCS was first studied from an information-theoretical perspective in~\cite{Verdu2002} by defining a metric called minimum energy per bit.
Then, the authors in~\cite{Prabhu2010} defined the energy-per-bit as the ratio of the consumed power to the achieved throughput expressed in J/bit.
Recently, the EE metric, defined as the ratio of the channel capacity $C$ to the power consumed to achieve this rate, was introduced in~[3], that is, $EE= \frac{C}{P_c+P_\textrm{Tx}}$,
where $P_{c}$ is the system's circuit power, and $P_\textrm{Tx}$ is the transmit power. Note that the spectral efficiency, $SE$ is defined as $SE= \frac{C}{B}$, where $B$ is the system bandwidth, meaning that  $EE= B \frac{SE}{P_c+P_\textrm{Tx}}$. In the rest of this paper, we adopt a unit bandwidth analysis where the $EE$ is expressed as $EE=  \frac{SE}{P_c+P_\textrm{Tx}}$ in bits/J.

Considering this relationship, it is clear that when SE is maximized for high values of $P_\textrm{Tx}$,  EE is reduced, thus the existence of a trade-off between SE and EE.

\subsection{Circuit Power Model}
\begin{figure}[h]
\begin{center}
\includegraphics[width=\figwidth]{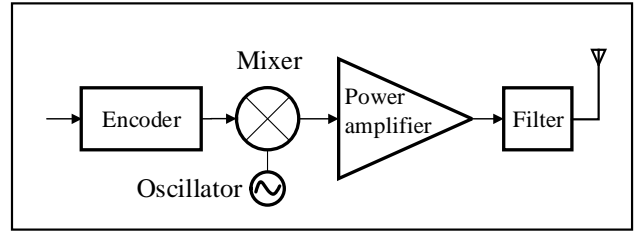}\\
(a)\\[.5cm]
\includegraphics[width=9cm]{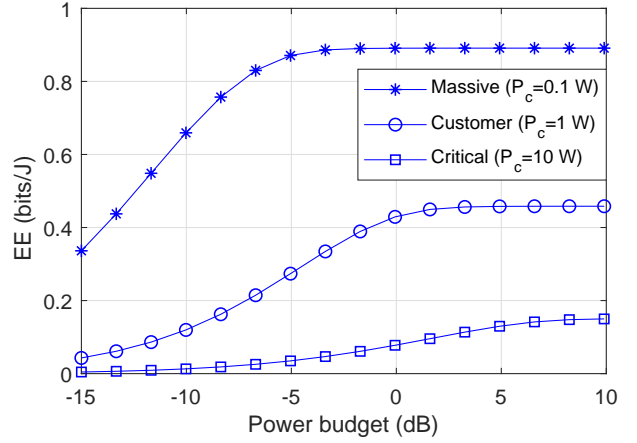}\\
(b)\\[.5cm]
\includegraphics[width=\figwidth]{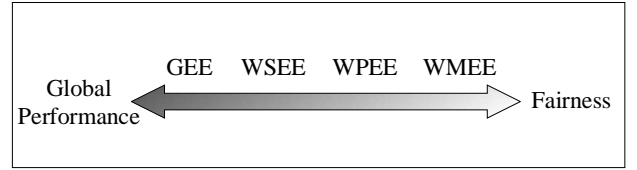}\\
(c)
\end{center}
\caption{(a) Circuit components involved in the circuit power; (b) various power profiles based on application type and circuit power value; (c) fairness degree of various different EE definitions\cite{Zappone2015}.}\label{fig_1_MAG}
\end{figure}

When defining the EE, the consumed power, which includes both the transmit power, $P_\textrm{Tx}$, and the circuit power, $P_{c}$, is required. The circuit power includes the power consumed by the communication chain (amplifiers, mixers, ADCs, DACs, filters, oscillators, etc) excluding the antenna as shown in Fig.~\ref{fig_1_MAG}.a. Note that the EE does not consider the overall power consumption of the WCS as it may be related to services other than the communication such as data processing and data storing for WSN, cooling for BSs, transportation for drones, etc.
In Fig.~\ref{fig_1_MAG}.b, we highlight the impact of $P_c$ on the EE in various applications, that is, massive, consumer, and critical applications, as a function of the available transmit power budget. We show, clearly that $P_c$ has a negative impact on EE since as $P_c$ increases, the EE is degrading.
\subsection{Energy Efficiency of Multi-Dimension WCS}
In the case of multi-dimension WCS (MD-WCS) such as multiple antennas, multiple subcarriers or multiple users systems, the EE can have multiple expressions, depending on the target criteria.
For example, in OFDM communications, if all subcarriers are equivalent, the employed EE expression is defined as the ratio of the total SE to the total power. In the case of an uplink cellular network, if the fairness is to be considered, EE could be defined as the product of partial EE achieved by each user. This way, we avoid situations where some users with bad channels may never communicate.

In the rest of this paper, $N$ denotes the dimension of the MD-WCS and  $i$ denotes the index of a given sub-system of~it.
As indicated in~\cite{Zappone2015}, there are mainly  four different ways to define the EE:
\subsubsection{Global EE (GEE)}~\\
The GEE is defined  as the ratio of global achievable rate to the total consumed power, that is, $GEE= \frac{\sum_i SE_i}{\sum_i P_{c,i}+P_{\textrm{Tx},i}}$. The GEE is simple to compute and characterizes the global performance of the WCS but does not allow sufficient fairness in MD-WCS. The GEE is wildly used in MIMO and OFDM applications where the fairness is not required among the antennas and the subcarriers, respectively.
\subsubsection{Weighted Sum EE (WSEE)}~\\
The WSEE is defined as the weighted sum of the partial EE's, that is, $WSEE= \sum_i \omega_i  \frac{SE_i}{P_{c,i}+P_{\textrm{Tx},i}}$ where $\omega_i$ is the weight associated to the {i-th} dimension. The  WSEE  can ensure a  priority to a subset of the system.
While the weights in the WSEE are chosen to highlight the importance of some parts of the MD-WCS, it does not guarantee total fairness.
The WSEE can be considered for heterogeneous networks (HetNets) applications where small cells have different weights than the macro cell when maximizing the EE.
\subsubsection{Weighted Product EE (WPEE)}~\\
The WPEE is defined as the weighted product of the partial EE's, that is, $WPEE= \prod_i \omega_i \frac{SE_i}{P_{c,i}+P_{\textrm{Tx},i}}$. The  WPEE aims to ensure higher fairness among all system's subsets.
The WPEE can be employed for cases of multiple users with no priority and with high fairness. The WPEE might be used in cellular applications, where the power and resource allocation at the cell maximize the EE, considering all users' EE. Otherwise, if one user is not connected the overall WPEE is reduced to zero.
\subsubsection{Weighted minimum of the energy efficiencies (WMEE)}~\\
The WMEE is defined as the minimum of the weighted   partial EE's, that is, $WMEE= \min_i \omega_i \frac{SE_i}{P_{c,i}+P_{\textrm{Tx},i}}$. The  WMEE can ensure higher fairness by focusing on maximizing the lowest partial EE among all system's subsets.
This EE is employed for cases of multiple users with a certain minimal EE per each user. An application of the WMEE is the case of WSN where a failure of one node may cause an outage of the network performance.

In Fig.~\ref{fig_1_MAG}.c, we present the different types of the EE and their corresponding degree of fairness. Note that the fairness is formally defined according to the Max-Min fairness. In other words, as this fairness increases, the distribution of the EE among the multiple communications links is more balanced.

\section{Energy Efficiency of SISO Systems}
SISO communications cover a wide range of applications such as WSN, mobile systems and public safety communications.
Usually, the transmit power of SISO systems is designed to maximize the SE under a certain power budget. The resulting power control is called the water-filling power allocation (WPA). While the WPA ensures obtaining the maximum SE, the corresponding EE is compromised which urges the need to develop a new power allocation scheme focusing on maximizing the EE.

\subsection{ SISO EE  Power Control}
Note that the EE function with respect to $P_\textrm{Tx}$ is not a convex but a pseudo-concave function, which is known to admit a unique maximizer~\cite{Zappone2015}. Therefore, many works in the literature used numerical methods to find the $P_\textrm{Tx}$ maximizing the EE~\cite{Isheden2012}. The most used method is the fraction programming method based on the Dinkelbach's algorithm~\cite{Zappone2015}. However, the reached solution is not entirely satisfactory as the lack of a closed-form solution does not allow to build sufficient insights into the problem. Recently, an explicit expression of the corresponding energy-efficient power allocation (EEPA) has been presented in~\cite{Sboui2015} as $P_{EE}(\gamma)=\frac{1}{\gamma} \left( {e}^{1+ W \left( \frac{\gamma P_c -1}{e} \right)}-1\right)$, where $\gamma$ is the channel realization modulus squared and $W$ is the main branch of W-Lambert function.
Interestingly, the EE expression, in~\cite{Sboui2015}, presented new insights on the EE performance and particularly on the EE-SE relationship.

\subsection{The SISO EE-SE  Relationship}
\begin{figure*}[h]
\begin{mdframed}[linewidth=.6pt]
  \centering
  \subcaptionbox{}{%
   \includegraphics[width=\figwidth]{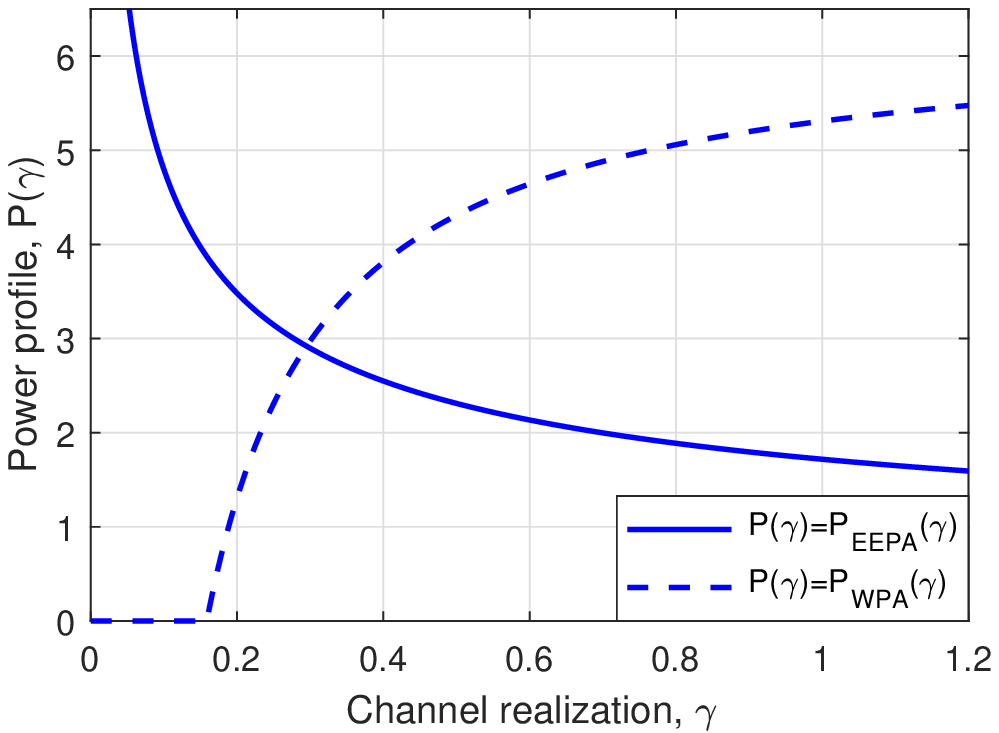}
   }\quad
  \subcaptionbox{}{%
   \includegraphics[width=\figwidth]{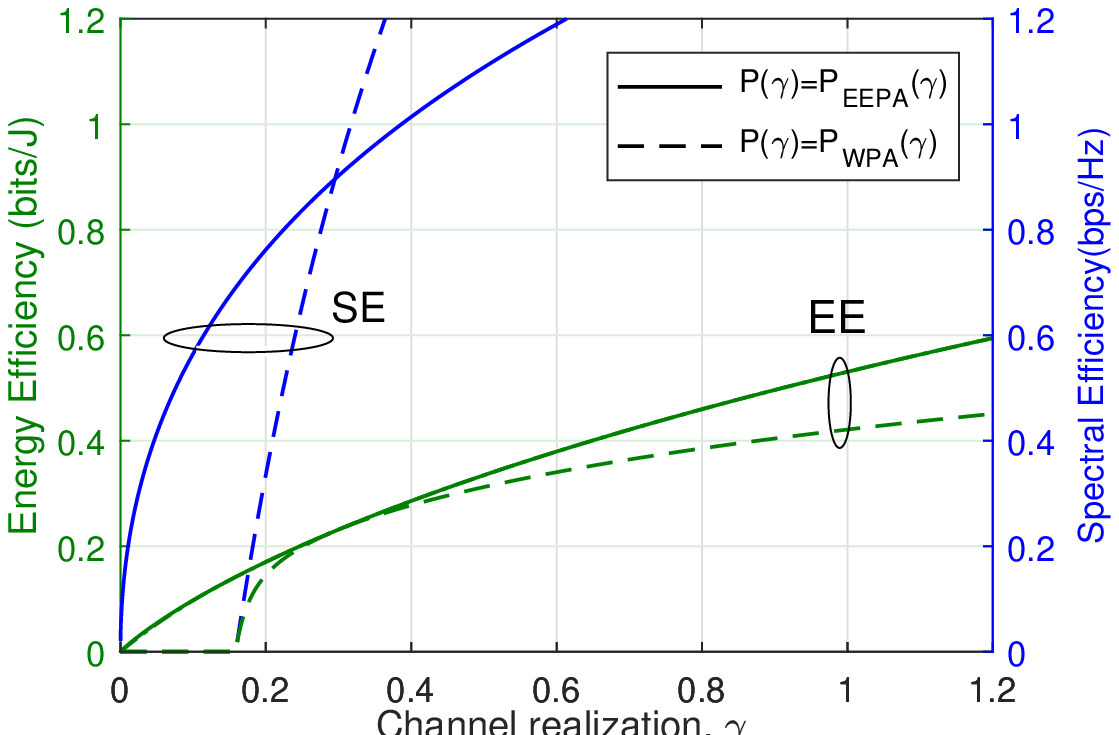}}
  \bigskip
  \subcaptionbox{}{%
    \includegraphics[width=\figwidth]{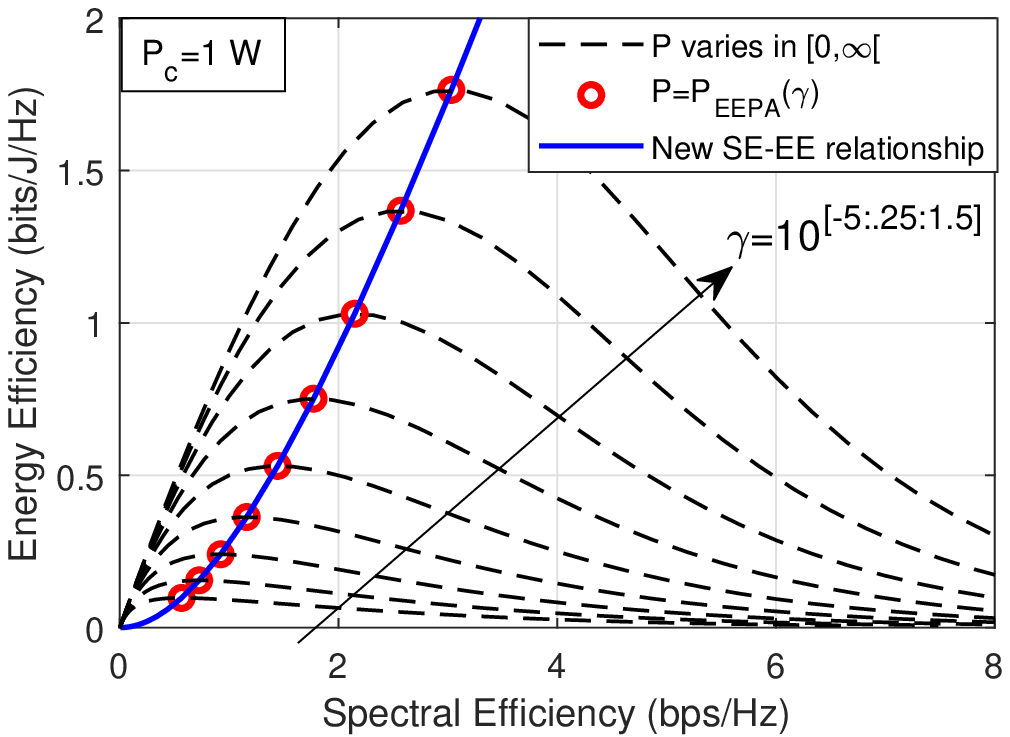}}\quad
  \subcaptionbox{}{%
    \includegraphics[width=\figwidth]{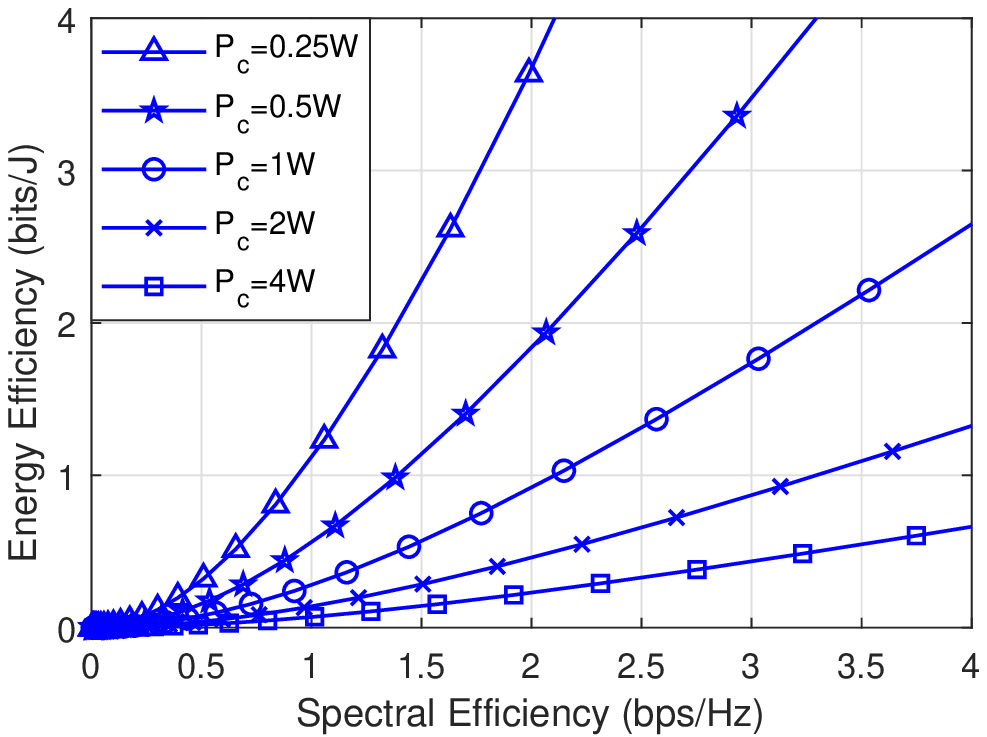}}
  \end{mdframed}
  \caption{(a) EE and SE based power profile~\cite{Sboui2017AccessEE}; (b) SE and the EE of both EE and SE based power control~\cite{Sboui2017AccessEE};\\ (c) The new EE and SE relationship in SISO communications~\cite{Sboui2015}; (d) The EE-SE relationship with different $P_c$.}\label{fig_MAG_EE_SISO_1}
\end{figure*}
In order to capture the mentioned new EE insights, we present in Fig.~\ref{fig_MAG_EE_SISO_1} the variation of the EE with the channel realization when an EEPA is employed and the corresponding  EE-SE variation.

In Fig.~\ref{fig_MAG_EE_SISO_1}.a, the EEPA, i.e., $P_{EE}(\gamma)$ is compared to the WPA, i.e., $P_{WPA}(\gamma)$. It is shown that both schemes behave oppositely as $\gamma$ increases. In fact, the WPA scheme is well-known to transmit when the channel is good and remains silent for bad channels in order to reach a high overall spectrum efficiency.
However, when the EE is maximized the EEPA scheme transmits with low power for good channels as the resulting SE divided by a low power corresponds to a high~EE.

In Fig.~\ref{fig_MAG_EE_SISO_1}.b, the EE and SE are presented for both EEPA and WPA schemes to have an idea of the resulting tradeoffs between maximizing the EE or SE. Interestingly, the SE of the EEPA is higher than the SE of the WPA for bad channels, in this case for $\gamma < 0.2$. However, the SE of the WPA overcomes the one of the EEPA which, gives in average, a maximum  SE.

This EEPA scheme is the essence of the new EE-SE relationship highlighted in Fig.~\ref{fig_MAG_EE_SISO_1}.c.
In this figure, we notice that the variation of the transmit power shows a pseudo concave EE function that has global maxima at a certain value of $P$. This value matches the optimal power maximizing the EE, i.e., $P_{EE}(\gamma)$ as $\gamma$ varies given by the red circles in the figure. Interestingly, the SE is increasing with the EE when the transmit power exceeds this value, the EE decreases and reaches zero. Interestingly, the set of the EE-SE  red circles, resulting from picking the $P_{EE}(\gamma)$ as $\gamma$ varies, presents a strictly increasing curve of the SE with respect to the EE. In other words, when the power is designed in a way to maximize the EE, the resulting SE is increasing with EE as shown by the blue curve in Fig.~\ref{fig_MAG_EE_SISO_1}.c for $P_c=1 W$

In Fig.~\ref{fig_MAG_EE_SISO_1}.d, we analyze the impact of the circuit power $P_c$ on the SISO performance. As shown in Fig.~\ref{fig_MAG_EE_SISO_1}.c, the SE computed using the power resulting from the EE maximization is increasing with the EE. However, the EE is decreasing by around $50\%$ as the $P_c$ doubles while conserving the same value of the SE. This observation shows the high influence of $P_c$ on the overall EE. In fact, one of the ways to considerably enhance the EE, while conserving the same data rate, is to design systems with low circuit power. Hence, efforts need to be made in the direction of reducing the circuit power while designing future wireless devices and equipments (sensors, mobile stations and base stations, etc.)
\begin{figure*}[t]
\begin{mdframed}[linewidth=.6pt]
  \centering
  \subcaptionbox{}{%
    \includegraphics[width=13cm]{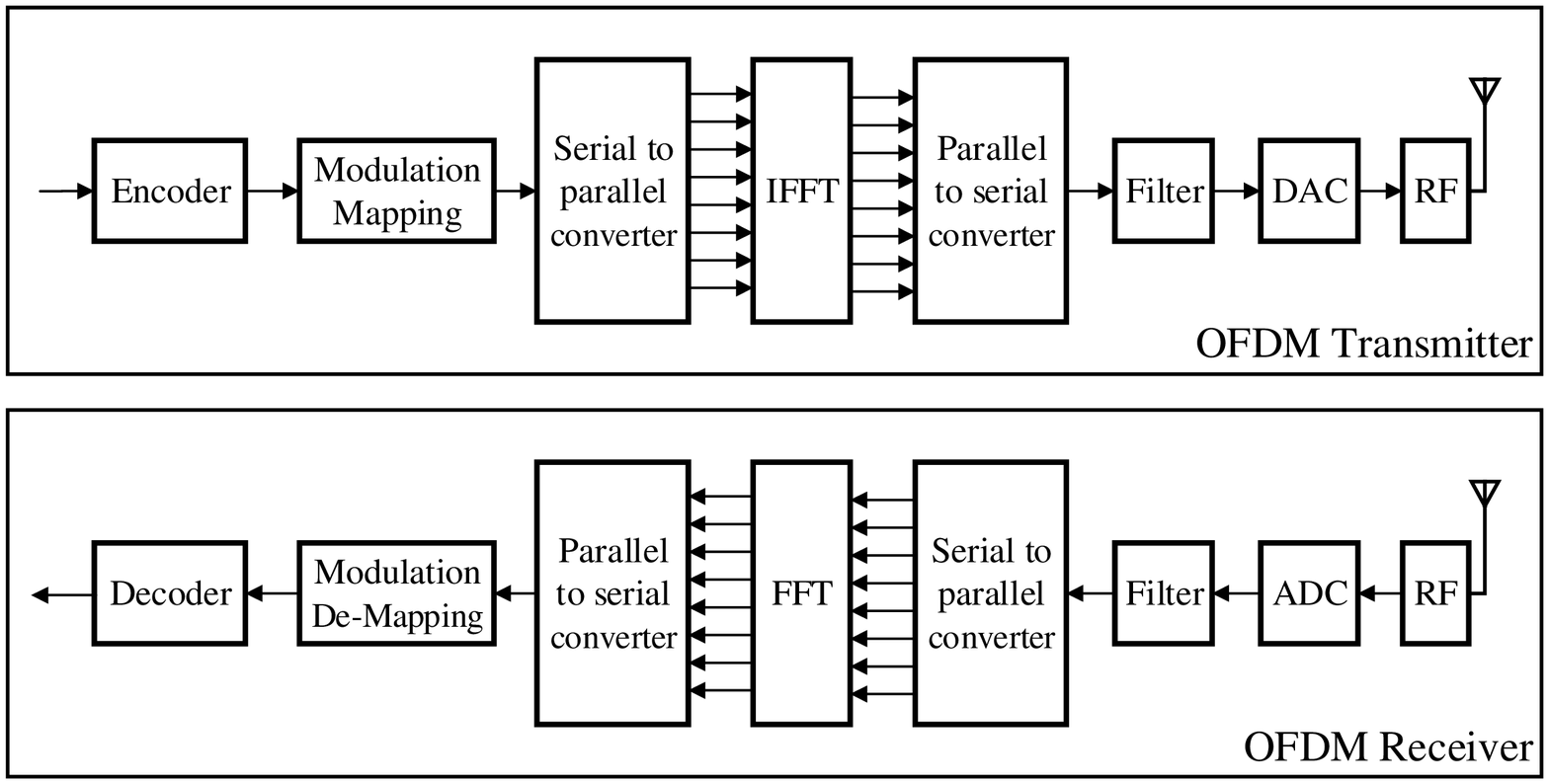}}
  \bigskip
  \subcaptionbox{}{%
    \includegraphics[width=\figwidth]{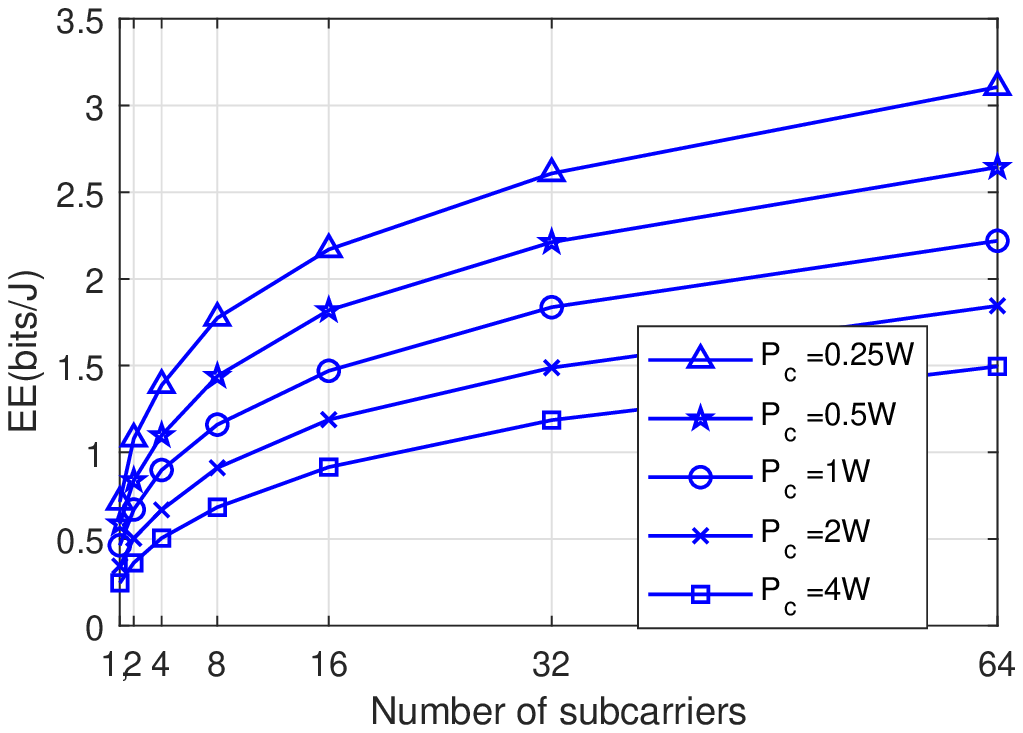}}\quad
  \subcaptionbox{}{%
    \includegraphics[width=\figwidth]{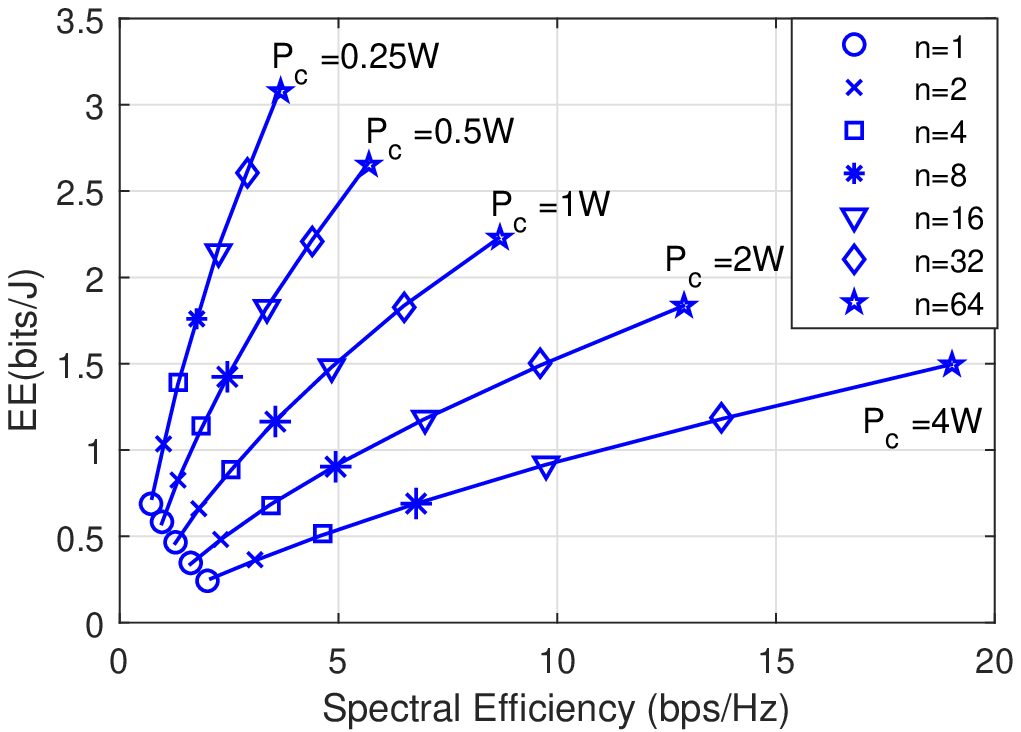}}
  \end{mdframed}
  \caption{(a) The OFDM communication chain; (b) EE variation with the number of subcarriers~\cite{Sboui2016PIMRC}; (c) The new EE and SE relationship in OFDM communications.}\label{fig_MAG_EE_OFDM_1}
\end{figure*}
\section{Energy Efficiency of  OFDM Systems}
\subsection{About the OFDM Technology}
The OFDM technique was introduced to avoid the channel variation and enhance the QoS by dividing the total bandwidth into $N$ subcarriers. 
This technique is based on transforming a serial wideband stream to narrowband parallel streams. The resulting signals are then transmitted on multiple orthogonal subcarriers that have different channel gains as shown in Fig.~\ref{fig_MAG_EE_OFDM_1}.a. The main advantage of OFDM is to overcome fading variations resulting in a robust and enhanced overall transmission used the correction codes~\cite{Li2011}.
The OFDM is currently part of many wireless communications protocols such as WiFi and WiMAX.

\subsection{OFDM EE Maximization}
Improving the EE of OFDM systems was performed, in previous works, based on minimizing the power consumption.
Then, the EE of the OFDM systems based on maximizing the EE metric was studied in~\cite{Venturino2015} and an explicit expression of the power was presented in~\cite{Sboui2016PIMRC} where the authors present analytical power allocation expressions in OFDMA cellular networks under unconstrained and constrained conditions regarding the power and rate requirements. These EEPA expressions, following the same insights of the SISO EEPA, present interesting results in terms of the EE of subcarriers scalability and the EE-SE relationship.

\subsection{OFDM EE-SE relationship}
In Fig.~\ref{fig_MAG_EE_OFDM_1}.b, we show that, for a given $P_c$, having more subcarriers, enhances the EE, meaning that considering an energy efficient transmission reduces the OPEX of cellular operators as the consumption is reduced for the same data rate. In addition, we notice that even if the circuit power increases, the slope of the EE curve is almost the same. This fact shows that the EE enhancement, when increasing the number of subcarriers, is highly predictable independently of the circuit power of the system.

A more interesting result related to the EE-SE relationship is presented in Fig.~\ref{fig_MAG_EE_OFDM_1}.c. We show that, for a fixed $P_c$, having more subcarriers enhances both the SE and the EE, which results again in an increasing SE with respect to the EE. Hence, the new EE-SE proportional relationship is also preserved in OFDM when EEPA schemes are adopted.

 \begin{figure*}[t]
\begin{mdframed}[linewidth=.6pt]
  \centering
  \subcaptionbox{}{%
    \includegraphics[width=10cm]{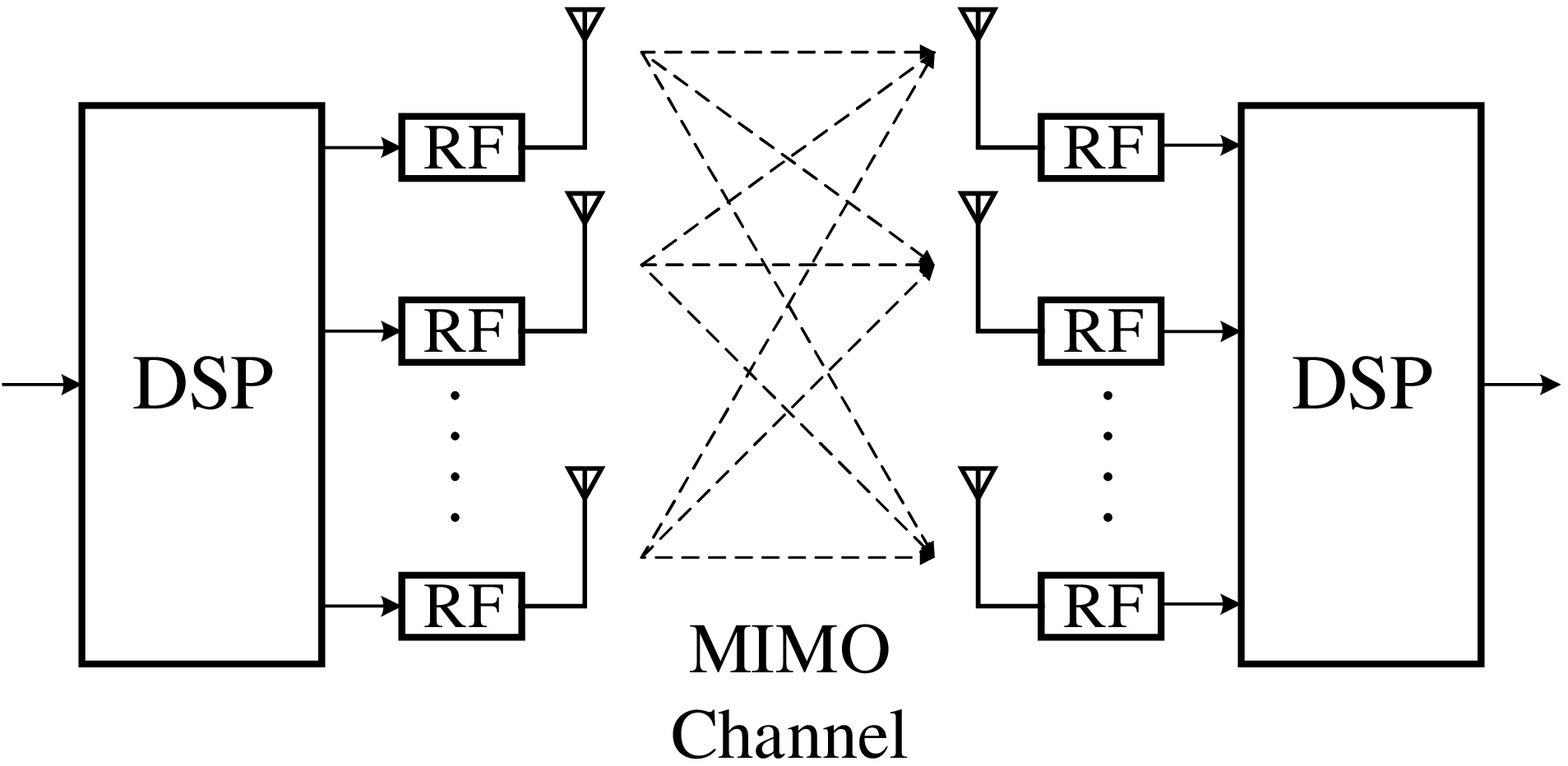}}\quad

  \bigskip
  \subcaptionbox{}{%
    \includegraphics[width=\figwidth]{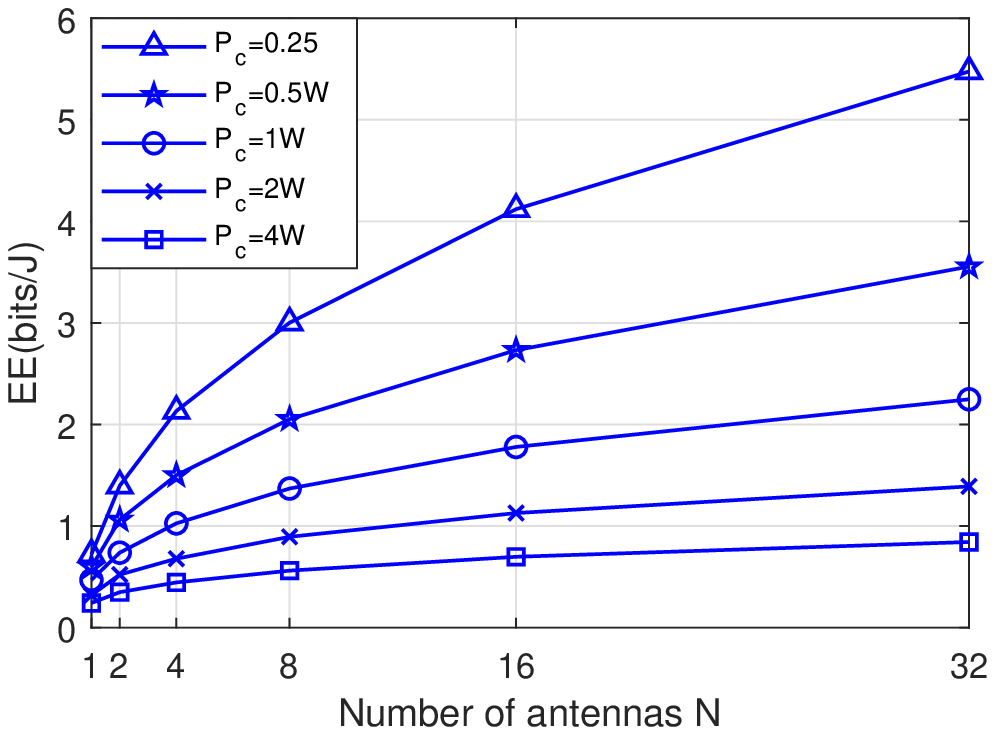}}\quad
  \subcaptionbox{}{%
    \includegraphics[width=\figwidth]{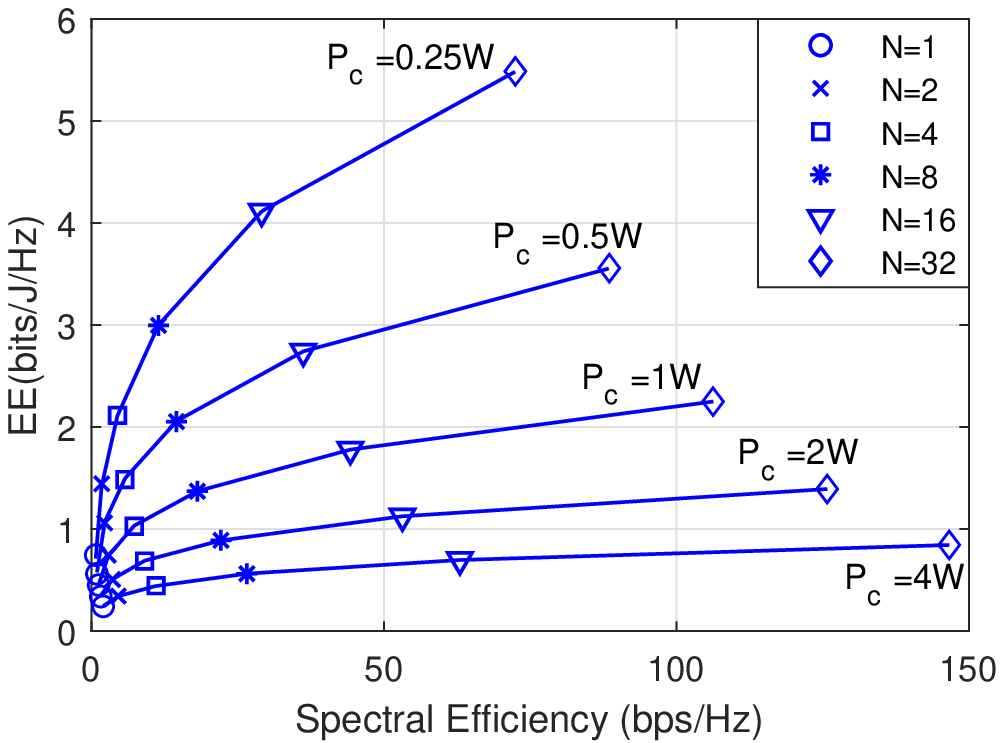}}
  \end{mdframed}
  \caption{(a) The MIMO communication chain; (b) EE variation with the number of antennas; (c) The new EE and SE relationship in MIMO communications.}\label{fig_MAG_EE_MIMO_1}
\end{figure*}
\section{Energy Efficiency of MIMO systems}
\subsection{About the MIMO Technology}
To improve the spectral efficiency, the MIMO concept was introduced in \cite{Telatar1999} as an alternative to reach higher data rate using the same bandwidth and power resources. The MIMO is a promising technology to remarkably increase the wireless channel capacity and allow to reach higher throughput. The MIMO is based on using multiple antennas at the transmitter and receiver as shown in Fig.~\ref{fig_MAG_EE_MIMO_1}.a.
The corresponding theoretical limit of the maximum data rate is proportional to the minimum between the transmit and receive antennas~\cite{Telatar1999} which presents a huge gain compared to SISO.

MIMO communications are based on two main techniques: multiplexing and beamforming. In multiplexing, the transmitter sends parallel streams of data at the same time in order to increase the overall data rate. In beamforming, the transmitter sends a narrow signals beam to the receiver and enhance the reliability.
MIMO technology is currently deployed in multiple IEEE 802.11n and the 3GPP LTE in the~4G.

\subsection{MIMO EE Maximization}
When the CSI is perfectly known, the singular value decomposition (SVD) of the channel matrix is employed in order to maximize the SE via leveraging parallel SISO channels. The MIMO-SVD is widely employed as it simplifies the power allocation procedure and results in high SE. Similarly, the EE of MIMO system can be maximized using the SVD~\cite{Zappone2015}.
The challenge of MIMO systems is again related to the non-convexity of the EE. Numerical fractional programming methods were presented as in~\cite{Zappone2015}, where the nonconvex problem is transformed into a subtractive one. Note that the EEPA of MIMO systems with the WSEE objective function is a natural extension of the SISO EEPA since each SVD parallel channel can be considered as a SISO system. However, in the case of the MIMO GEE objective function, a new scheme was presented in \cite{Sboui2017AccessEE} with the closed-form expression of the optimal power derived under certain conditions. As we mentioned before, the WPEE and WMEE are not widely used for MIMO systems since there is no fairness requirement among the different antennas.

\subsection{MIMO EE-SE relationship}
We first start, in Fig.~\ref{fig_MAG_EE_MIMO_1}.b, by exploring the effect of increasing the number of antennas $N$ on the EE with an EEPA scheme. We show that the EE increase is logarithmic with $N$ which gives a general insight about the EE gain of MIMO systems compared to the SISO systems. In particular, as the circuit power is lower, the gain in EE is higher as $N$ increases.

From another side, in Fig.~\ref{fig_MAG_EE_MIMO_1}.c, we analyze the EE-SE relationship when the number of antennas $N$ increases. We confirm again, in MIMO, that the EE-SE relationship is proportional, since an increasing $N$ results in higher EE and SE. In addition, we highlight the effect of the circuit power $P_c$ on the EE-SE relationship. For the same SE, as $P_c$ decreases, EE is increasing remarkably. For instance, for $N=4$, the MIMO system is twice more energy efficient with $P_c=0.25W$ than when $P_c=1W$. Which means, in this case, that mobile devices will have a double lifetime, or that mobile operators will pay half the electricity bill.

\section{Results Discussion and Related Research Problems}
We present in this part a discussion about the aforementioned results as well as suggestion of related research problems.
\subsection{Results Discussion}
After our analysis of the EE-SE relationship, we clearly show that the EE is an increasing function of SE for all technologies. In addition, in order to highlight the differences and similarities in SISO, OFDM, and MIMO, we present in Table~\ref{tab1} a summary of the EE and SE performance results. In particular, in MD-WCS, as we increase the dimension~$N$, the system becomes more energy efficient in addition to providing higher spectral efficiency. As a result, when OFDM is combined with MIMO, we expect to have even higher EE and SE. For example, for $P_c=1$W, the OFDM EE increases by more than three folds from~1 to~16 subcarriers (with one antenna) where as the MIMO EE is more than the double for~4 antennas compared to SISO (with on subcarrier). As a conclusion, the MIMO-OFDM EE, with~16 subcarriers and~4 antennas, is expected to reach more than six times the SISO EE. This observation reflects a high synergy between OFDM and MIMO in WCS which is already observed in terms of SE in high speed technologies such as WiMAX LTE, LTE-A, and IEEE 802.11n/ac/ax.
{\linespread{1}
\begin{table}
\centering
\caption{\label{tab1}Analysis summary for SISO, OFDM, and MIMO.}
\addtolength{\tabcolsep}{-3pt}
{\renewcommand{\arraystretch}{2.5}

\begin{tabular}{|c|c|c|c|c|c|c|}
\hline
\textbf{WCS}&\textbf{\pbox{2.5cm}{Detailed\\power\\profile}}&\textbf{\pbox{1.7cm}{EE shape in\\MD-WCS}}&\textbf{\pbox{1.5cm}{EE-SE\\shape}}&\textbf{\pbox{1.5cm}{EE gain\\compared\\to SISO$^*$}}&\textbf{\pbox{1.5cm}{SE gain\\compared\\to SISO$^*$}}\\[2mm]   \hline \hline
\textbf{ SISO }            & \cite{Sboui2015}          &    -                          & \pbox{1.5cm}{increasing, \\parabolic  }           & -     &        -   \\[1mm] \hline
\textbf{ OFDM }            &  \cite{Sboui2016PIMRC}      &   \pbox{1.5cm}{increasing, \\logarithmic} & \pbox{1.5cm}{increasing,\\linear} & \pbox{1.5cm}{3x (N=16)\\5x (N=64)} & \pbox{1.5cm}{4x (N=16)\\7x (N=64)}          \\[1mm] \hline
\textbf{ MIMO }  &  \cite{Sboui2017AccessEE}   &   \pbox{1.5cm}{increasing, \\logarithmic}  & \pbox{1.5cm}{increasing,\\logarithmic} & \pbox{1.5cm}{2x (N=4)\\5x (N=32)}  & \pbox{1.5cm}{6x (N=4)\\84x (N=32)} \\[1mm] \hline
\end{tabular}
}{
\begin{flushleft}
$^*$: for $P_c= 1$W.
\end{flushleft}
}\vspace{-7mm}
\end{table}
 }
As a result, the new EE-SE relationship has an important impact on the design of 5G systems from two perspectives. First, the transmit power profile is expected to be radically changed from the traditional WPA to the EEPA described in  Fig.~\ref{fig_MAG_EE_SISO_1}.a. Second, the EE is expected to be the primary performance metric instead of the SE. In fact, EE considerations are mandatory for energy-sensitive applications in order to extend operation time, and is recommended in other applications to reduce operation costs.
\subsection{Related Research Problems}
We present some interesting research problems related to the EE requiring future investigations.
\begin{itemize}
\item \textbf{MIMO with CSI Unavailability:}
In some practical MIMO scenarios, the CSI is not available or imperfect. Hence, the EEPA is hard to obtain and further algorithms to find the maximal EE with imperfect CSI need to be developed. One possible method is to maximize an approximation of an average EE considering the channel gain statistics.
 \item \textbf{Multi-Users:}
In 4G-LTE and 5G networks, there are multiple users. The corresponding  EE problem is interesting to study. In fact, each user is trying to maximize its EE compared to the EE maximization of the global system. Hence, the WMEE objective function can be studied.
 \item \textbf{Multi-Cells:}
Another interesting case would be the multi-cell scenario where interference from other cells is
considered. The case of rate constraints where a minimum rate (as a measure of QoS) for
each user must be guaranteed would be of interest.
 \item \textbf{Massive MIMO:}
Massive MIMO is considered as an important part of the coming 5G networks. Hence, studying its EE is relevant as in massive MIMO, the consumption of the computations required for computing the precoder and decoder is relatively huge and should be included in the EE objective function.
Also, the proposed method requires channel state information. The overhead (spectrum efficiency loss) for obtaining such a channel state information increases with N, thus impacting the EE.
\item \textbf{HetNets:}
A heterogeneous network consists of multiple nodes that differ in radio access interfaces, transmit power, and coverage range. Since HetNets are considered as an enabling technology of the 5G which promotes more energy efficient communications, it is interesting to study the implications of utilizing several radio interfaces on the energy efficiency in multi-radio IoT/user applications.
\end{itemize}

%
%
\section{Conclusion}
In this paper, we presented a novel perspective regarding the EE-SE relationship. This new relationship is based on an increasing EE with SE. We presented multiple varieties of the EE metric and studied the corresponding performance for SISO, OFDM, and MIMO communications. Our study is based on a new power control scheme maximizing the EE rather than the SE. We highlighted that not only does the proposed scheme improve the EE, but it shows a new EE-SE proportional relationship. In addition, we highlighted the impact of this new relationship on the design of the future 5G systems. Finally, we presented some interesting research topics related to the EE.

\bibliographystyle{IEEEtran}
\bibliography{References_Mag}

\begin{thebibliography}{10}
\providecommand{\url}[1]{#1}
\csname url@samestyle\endcsname
\providecommand{\newblock}{\relax}
\providecommand{\bibinfo}[2]{#2}
\providecommand{\BIBentrySTDinterwordspacing}{\spaceskip=0pt\relax}
\providecommand{\BIBentryALTinterwordstretchfactor}{4}
\providecommand{\BIBentryALTinterwordspacing}{\spaceskip=\fontdimen2\font plus
\BIBentryALTinterwordstretchfactor\fontdimen3\font minus
  \fontdimen4\font\relax}
\providecommand{\BIBforeignlanguage}[2]{{%
\expandafter\ifx\csname l@#1\endcsname\relax
\typeout{** WARNING: IEEEtran.bst: No hyphenation pattern has been}%
\typeout{** loaded for the language `#1'. Using the pattern for}%
\typeout{** the default language instead.}%
\else
\language=\csname l@#1\endcsname
\fi
#2}}
\providecommand{\BIBdecl}{\relax}
\BIBdecl

\bibitem{ITUreport2015}
\phantom{ } ITU-R Rec.~M.2083, ``{IMT} vision {–} framework and overall
  objectives of the future development of {IMT} for 2020 and beyond,'' Tech.
  Rep.~4, Oct. 2015.

\bibitem{Xu2015}
J.~Xu, L.~Duan, and R.~Zhang, ``Cost-aware green cellular networks with energy
  and communication cooperation,'' \emph{IEEE Communications Magazine},
  vol.~53, no.~5, pp. 257--263, May 2015.

\bibitem{Li2011}
G.~Li, Z.~Xu, C.~Xiong, C.~Yang, S.~Zhang, Y.~Chen, and S.~Xu,
  ``Energy-efficient wireless communications: {T}utorial, survey, and open
  issues,'' \emph{IEEE Wireless Communications Magazine}, vol.~18, no.~6, pp.
  28--35, Dec. 2011.

\bibitem{Cover2006}
T.~M. Cover and J.~A. Thomas, \emph{Elements of Information Theory},
  2nd~ed.\hskip 1em plus 0.5em minus 0.4em\relax Wiley-Interscience, 2006.

\bibitem{Han2011}
C.~Han, T.~Harrold, S.~Armour, I.~Krikidis, S.~Videv, P.~M. Grant, H.~Haas,
  J.~S. Thompson, I.~Ku, C.~X. Wang, T.~A. Le, M.~R. Nakhai, J.~Zhang, and
  L.~Hanzo, ``Green radio: radio techniques to enable energy-efficient wireless
  networks,'' \emph{IEEE Communications Magazine}, vol.~49, no.~6, pp. 46--54,
  June 2011.

\bibitem{Guo2014}
Y.~Guo, J.~Xu, L.~Duan, and R.~Zhang, ``Joint energy and spectrum cooperation
  for cellular communication systems,'' \emph{IEEE Transactions on
  Communications}, vol.~62, no.~10, pp. 3678--3691, Oct. 2014.

\bibitem{Sboui2015}
L.~Sboui, Z.~Rezki, and M.-S. Alouini, ``Energy-efficient power allocation for
  underlay cognitive radio systems,'' \emph{IEEE Transactions on Cognitive
  Communications and Networking}, vol.~1, no.~3, pp. 273--283, Sept. 2015.

\bibitem{Verdu2002}
S.~Verdu, ``Spectral efficiency in the wideband regime,'' \emph{IEEE
  Transactions on Information Theory}, vol.~48, no.~6, pp. 1319--1343, June
  2002.

\bibitem{Prabhu2010}
R.~S. Prabhu and B.~Daneshrad, ``Energy-efficient power loading for a
  {MIMO-SVD} system and its performance in flat fading,'' in \emph{Proc. of the
  IEEE Global Telecommunications Conference, GLOBECOM'10, Miami, Florida, USA},
  Dec. 2010, pp. 1--5.

\bibitem{Zappone2015}
A.~Zappone and E.~Jorswieck, ``Energy efficiency in wireless networks via
  fractional programming theory,'' \emph{Foundations and Trends in
  Communications and Information Theory}, vol.~11, no. 3-4, pp. 185--396, 2015.

\bibitem{Isheden2012}
C.~Isheden, Z.~Chong, E.~Jorswieck, and G.~Fettweis, ``Framework for link-level
  energy efficiency optimization with informed transmitter,'' \emph{IEEE
  Transactions on Wireless Communications}, vol.~11, no.~8, pp. 2946--2957,
  2012.

\bibitem{Sboui2017AccessEE}
L.~Sboui, Z.~Rezki, and M.~Alouini, ``Energy-efficient power allocation for
  {MIMO-SVD} systems,'' \emph{IEEE Access}, vol.~5, pp. 9774--9784, 2017.

\bibitem{Sboui2016PIMRC}
L.~Sboui, Z.~Rezki, and M.~S. Alouini, ``Energy-efficient power control for
  {OFDMA} cellular networks,'' in \emph{In Proc. of the IEEE 27th Annual
  International Symposium on Personal, Indoor, and Mobile Radio Communications
  (PIMRC'16), Valencia, Spain}, Sept. 2016, pp. 1--6.

\bibitem{Venturino2015}
L.~Venturino, A.~Zappone, C.~Risi, and S.~Buzzi, ``Energy-efficient scheduling
  and power allocation in downlink {OFDMA} networks with base station
  coordination,'' \emph{IEEE Transactions on Wireless Communications}, vol.~14,
  no.~1, pp. 1--14, Jan. 2015.

\bibitem{Telatar1999}
\BIBentryALTinterwordspacing
E.~Telatar, ``Capacity of multi-antenna {G}aussian channels,'' \emph{European
  Transactions on Telecommunications}, vol.~10, no.~6, pp. 585--595, 1999.
  [Online]. Available: \url{http://dx.doi.org/10.1002/ett.4460100604}
\BIBentrySTDinterwordspacing

\end{thebibliography}

\end{document}